# Pseudogap isotope effect as a probe of bipolaron mechanism in high temperature superconductors

Victor D. Lakhno

Keldysh Institute of Applied Mathematics of Russian Academy of Sciences,
125047 Moscow, Russia; lak@impb.ru

**Abstract:** A theory of a pseudogap phase of high-temperature superconductors where current carriers are translation invariant bipolarons is developed. A temperature $T^*$ of a transition from a pseudogap phase to a normal one is calculated. For the temperature of a transition to the pseudogap phase, the isotope coefficient is found. It is shown that the results obtained, in particular, the possibility of negative values of the isotope coefficient are consistent with the experiment. New experiments on the influence of the magnetic field on the isotope coefficient are proposed.

**Keywords:** incoherent electron pairs; Pekar-Fröhlich Hamiltonian; charged Bose gas; optical phonon

**1. Introduction**

Among the most amazing and mysterious phenomena of high-temperature superconductivity (HTSC) is the existence of a pseudogap phase at a temperature above the critical temperature of a superconducting (SC) transition [1–5]. In a pseudogap phase the spectral density of states near the Fermi surface demonstrates a gap for $T > T_c$, where $T_c$ is a temperature of a SC transition which persists up to the temperatures $T^*(T^* > T_c)$ above which the pseudogap disappears. Presently the explanation of this phenomenon is reduced to two possibilities. According to the first one, it is believed that for $T > T_c$ some incoherent electron pairs persist in the sample, while for $T < T_c$ their motion becomes coherent and they pass on to the SC state. For $T > T^*$, the pairs disintegrate and the pseudogap state disappears [6–9]. According to the second one, the transition to the pseudogap phase is not concerned with superconductivity, but is caused by the formation of a certain phase with a hidden order parameter or a phase with spin fluctuations [10–12].

Presently the first viewpoint on the nature of a pseudogap in HTSC increasingly dominates which is associated with the idea that paired electron states exist for $T > T_c$. The question of the nature of paired electron states per se remains open. In this paper paired electron states are taken to be translation invariant (TI) bipolarons.

The TI bipolaron theory of SC based on the Pekar-Fröhlich Hamiltonian of electron-phonon interaction (EPI) when EPI cannot be considered to be weak as distinct from the Bardeen–Cooper–Schrieffer theory [12], was developed in papers [13–15] (see also review [16]). The role of Cooper pairs in this theory belongs to TI bipolarons whose size ($\approx 1nm$) is much less than that of Cooper pairs ($\approx 10^3 nm$). According to [13–16], in HTSC materials TI bipolarons are formed near the Fermi surface and represent a charged Bose gas capable of experiencing Bose-Einstein condensation (BEC) at high critical temperature which determines the temperature of a SC transition.

As distinct from Cooper pairs, TI bipolarons have their own excitation spectrum [13–16]:

$$E_k^{bp} = E_{bp}\Delta_{k,0} + (\omega_0 + E_{bp} + k^2/2M_e)(1 - \Delta_{k,0}), \qquad (1)$$
$$M_e = 2m, \ \Delta_{k,0} = 1 \text{ for } k=0 \text{ and } \Delta_{k,0} = 0 \text{ for } k \neq 0,$$

where $E_{bp}$ is the ground state energy of a TI bipolaron (reckoned from the Fermi level), $\omega_0$ is the frequency of an optical phonon ($\hbar = 1$), $m$ is a mass of a band electron (hole), $k$ is a wave vector numbering excited states of a TI bipolaron.

This spectrum has a gap which in the isotropic case is equal to the frequency of an optical phonon $\omega_0$. At that, the inequality $\omega_0 \gg |E_{bp}|$ corresponds to the case of a weak EPI, $\omega_0 \ll |E_{bp}|$ - to the case of strong coupling and $\omega_0 \sim |E_{bp}|$ - to the case of intermediate coupling. According to [13]-[16], the number of TI bipolarons $N_{bp}$ at temperature T=0 is equal to: $N_{bp} \cong N\omega_0/2E_F$, where $N$ is the total number of electrons (holes), $E_F$ is the Fermi energy, i.e. $N_{bp} \ll N$.

The scenario of a SC based on the idea of a TI bipolaron as a fundamental boson responsible for superconducting properties explains many thermodynamic and spectroscopic properties of HTSC [13–16]. For this reason, the problem of the temperature of a transition to the pseudogap state and its isotope dependence is of interest.

## 2. The critical temperature of a pseudogap phase

Obviously, the temperature of a transition from a pseudogap phase to a normal one $T^*$ in this model is determined by disintegration of TI bipolarons into individual TI polarons. Thermodynamically, the value of $T^*$ should be determined from the condition that the free energy of a TI bipolaron gas exceeds the free energy of a TI polaron gas determined by the spectrum of TI polarons:

$$E_k^P = E_p \Delta_{k,0} + (\omega_0 + E_p + k^2/2m)(1 - \Delta_{k,0}), \qquad (2)$$

where $E_p$ is the energy of the polaron ground state.

For further consideration it is significant that the number of TI bipolarons in HTSC compounds is $N_{bp} \ll N$. For $n = N/V = 10^{21} cm^{-3}$, where $V$ is the system volume, the typical values of $n_{bp} = N_{bp}/V$ are of the order of $n_{bp} \sim 10^{18} - 10^{19} cm^{-3}$ [13]-[16]. Taking into account that $T^* > T_c$, in order to calculate the statistical sum of the bipolaron gas $Z_{bp}$ in the vicinity of $T^*$ one can use a classical approximation which requires that in the region of stability of the bipolaron gas the inequality:

$$T^* > T > T_c \qquad (3)$$

be fulfilled. In this case the expression for the statistical sum of the TI bipolaron gas has the form:

$$Z_{bp} = \frac{1}{h^{3N_{bp}} N_{bp}!} \prod_{i=1}^{N_{bp}} \int d^3 k_i e^{-E_{k_i}^{bp}/T}$$

$$= \left[ e^{-(\omega_0 + E_{bp})/T} \left( \frac{2\pi M_e T}{h^2} \right)^{3/2} \frac{eV}{N_{bp}} \right]^{N_{bp}}, \qquad (4)$$

where $e \approx 2.718$ is the natural logarithm base, $h = 2\pi\hbar$ is Planck constant (we use the energy units and put Boltzmann constant to unity).

Accordingly, for the statistical sum of a TI polaron gas formed as a result of disintegration of TI bipolarons, similarly to (4), we get:

$$Z_p = \left[ e^{-(\omega_0 + E_p)/T} \frac{(2\pi m T)^{3/2}}{h^3} \frac{eV}{2N_{bp}} \right]^{2N_{bp}} \qquad (5)$$

The condition of stability of a TI bipolaron gas with respect to its decay into a TI polaron gas is written as:

$$Z_{bp} \geq Z_p, \qquad (6)$$

where the equality describes the case of an equilibrium between the two gases which corresponds to the equation for temperature $T^*$ of a transition from a normal phase to a pseudogap one.
Substitution of (4), (5) into (6) leads to the condition:

$$\Delta = |E_{bp}| + \omega_0 - 2|E_p| \geq \frac{3}{2} T \ln æT, \qquad æ = \left(\frac{e}{4}\right)^{\frac{2}{3}} \frac{\pi m}{n_{bp}^{\frac{2}{3}} h^2} \qquad (7)$$

In the case of equality expression (7) yields the equation for determining $T^*$:
$$z = We^W, \qquad T^* = æ^{-1}e^W, \qquad z = 2æ\Delta/3 \qquad (8)$$

Figure 1 shows the solution W=W(z) (Lambert function) (8) on condition that requirement (3) is fulfilled.

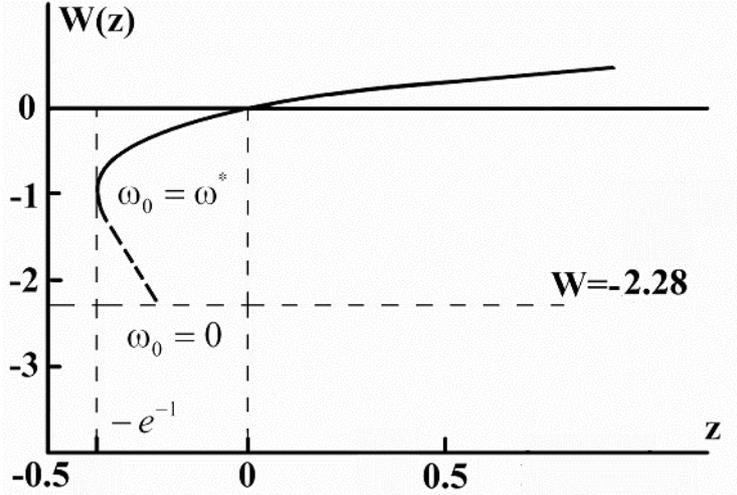

**Figure 1.** Solution W(z) of equation (8).

It holds on the interval $-e^{-1} < z < \infty$. On the interval $-e^{-1} < z < 0$ Lambert function is negative. Requirement (3) leads to the condition: $-2.28 < W < \infty$. Taking into account the expression for the temperature of a SC transition obtained in [13]-[16]:

$$T_c = T_c(\omega_0) = \left(F_{3/2}(0)/F_{3/2}(\omega_0/T_c)\right)^{3/2} T_c(0), \qquad (9)$$

$$T_c(0) = 3{,}31\hbar^2 n_{bp}^{2/3}/M_e, \quad F_{3/2}(x) = \frac{2}{\sqrt{\pi}} \int_0^\infty \frac{t^{1/2} dt}{e^{t+x} - 1},$$

we express $T^*$ as:
$$T^* \approx 9{,}8 \left(F_{3/2}(\omega_0/T_c)/F_{3/2}(0)\right)^{2/3} T_c \exp W \qquad (10)$$

Thus for example, for $\omega_0 \approx T_c$ we obtain from (10) $T^* \approx 3T_c(1)\exp W$, where $T_c(1)$ is determined by (9): $T_c(1) \approx 3.3T_c(0)$ that is for W=0 the pseudogap temperature $T^*$ exceeds the temperature of a SC transition $T_c$ more than threefold.

For $\omega_0 \gg T_c$ the temperature of a pseudogap phase is $T^* \gg T_c(1)$. In this case for estimating $T^*$ we can use an approximate formula derived from (8):

$$T^* \approx \frac{2}{3}\Delta/\ln\frac{2}{3}æ\Delta, \quad æ|\Delta| > \frac{3}{2} \qquad (11)$$

This limit, however, is observed in experiments only rarely. It can be concluded that in HTSC materials the main contribution into EPI leading to SC is made by phonon frequencies with $\omega_0 < T_c$ ($\omega_0 < 10\ meV$). This estimate is an order of magnitude less than the estimates of phonon frequencies which are generally believed to make the main contribution into the SC [13]. It also follows from (11) that $T^*$ grows only logarithmically as the concentration of $n_{bp}$ increases, while $T_c \sim n_{bp}^{2/3}$. Hence for a certain value of $n_{bp}$, the condition $T^* < T_c$ can be fulfilled which corresponds to disappearance of a pseudogap phase as it follows from the form of the exact solution of equation (8). In HTSC materials this takes place as doping increases to an optimal value for which the pseudogap phase no longer exists.

Notice that the obtained results derived by comparison of bipolaron and polaron classical canonical partition functions are not specific for the bipolarons. For example, the analog of expression (11) was earlier obtained in [17] using Saha equation for free fermion and bosonic molecular gases.

## 3. Isotope Coefficient for the Pseudogap Phase

The TI bipolaron theory of the pseudogap phase developed above enables one to investigate its isotopic properties.

It follows from (8) that like the SC phase, the pseudogap one possesses the isotopic effect. According to (8), the isotope coefficient is:

$$\alpha^* = -dlnT^*/dlnM, \qquad (12)$$

where $M$ is the mass of an atom replaced by its isotope. With regard to the fact that $\omega_0 \sim M^{-1/2}$ it takes the form (see Appendix):

$$\alpha^* = \frac{\omega_0}{3T^*} \frac{1}{1 + W(z)} \qquad (13)$$

It follows from (13) and Figure 1 that for the lower branch: $W(z) < -1$ and $\alpha^* < 0$. Accordingly, for the upper branch: $W(z) > -1$ and $\alpha^* > 0$. It should be noted that the upper branch corresponds to $\omega_0 > \omega^*$, while the lower branch to $\omega_0 < \omega^*$, where $\omega^* \approx T_c$. Expressions (10), (13) yield:

$$W = ln\left[c\left(F_{3/2}(0)/F_{3/2}(\omega_0/T_c)\right)^{2/3} T^*/T_c\right], \qquad (14)$$

where $c \approx 0.1$. Figure 2 illustrates the graph of the dependence $\alpha^*(\omega_0)$.

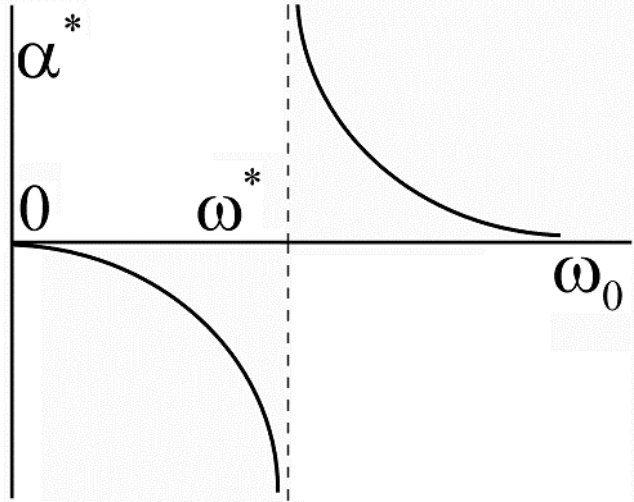

**Figure 2.** Dependence of the isotope coefficient for the pseudogap temperature $T^*$ on the phonon frequency $\omega_0 (\omega^* \approx T_c)$.

Hence, depending on the value of the phonon frequency $\omega_0$ coefficient $\alpha^*$ can have any sign and the value: $\alpha^* < 0$ for $\omega_0 < \omega^*$ and $\alpha^* > 0$ for $\omega_0 > \omega^*$. For $\omega_0 = \omega^*$, the isotope coefficient becomes infinite: $\alpha^*(\omega^* \pm 0) = \pm\infty$. The results obtained suggest that the isotope exponent diverges for $\omega_0 \to \omega^*$, i.e. for $T^* \approx T_c$ (Figure 2). Great negative values of the isotope coefficient in the pseudogap state were observed experimentally in [18]-[20]. It should be noted that in some cases negative values of the isotope coefficient were also observed in ordinary SC [21], which exceeded in modulus the value of the isotope coefficient in monoatomic systems α=0.5 yielded by the BCS. According to the theory suggested this is possible for $T^* \approx T_c$.

## 4. Isotope coefficient for pseudogap phase in magnetic field

According to [15], the spectrum of a TI bipolaron in a magnetic field $\vec{B}$ is determined by the modified expression (1):

$$E_k^{\text{Bp}} = E_{\text{Bp}}\Delta_{k,0} + (\omega_0 + E_{\text{Bp}} + k^2/2M_e + \mathrm{n}_{\text{Bp}}(\vec{B}\vec{k})/2M_e)(1 - \Delta_{k,0}) \qquad (15)$$

where the quantity η$_{бр}$, according to [15], is related to the first critical field SC $B_{max}$ as:

$$\eta_{бр} = \sqrt{\frac{2\omega_0 M_e}{B_{max}}}$$

The spectrum of a TI polaron will be determined by relation (2), since in weak fields $B < B_{max}$ the magnetic field leaves it practically unchanged ($B_{max}$, being the first critical field, is always much less than the value of the quantizing magnetic field of a TI polaron).

Performing calculations similar to Section 2, we obtain the same relations as in Sections 2, 3 where $\omega_0$ should be replaced by $\widetilde{\omega}_0 = \omega_0(1 + B^2/B_{max}^2)$. Thus, for example, the graph of the dependence $\alpha(\omega_0)$, determined by Fig.2, in a magnetic field will be a graph of the dependence $\alpha(\widetilde{\omega}_0)$. Hence, for $\widetilde{\omega}_0 < \omega^*$ an increase in the field value will lead to larger negative values of the isotope coefficient, and for $\widetilde{\omega}_0 > \omega^*$ to smaller positive values $\alpha^*$. In particular, a situation is possible when, for a certain value of the field, the isotope coefficient, being negative, as the field increases, becomes infinite at the point $\widetilde{\omega}_0 = \omega^*$ and becomes positive for $\widetilde{\omega}_0 > \omega^*$.

It will be interesting to verify these conclusions experimentally.

## 5. Discussion

The approach developed is based on the idea that in such HTSC as MgB$_2$, Bi-2223, Bi-2212, YBaCuO etc. at temperatures substantially above critical temperature $T_c$ electron pairs in the form of TI-bipolarons represent the noninteracting strongly bound bosons, demonstrating with decreasing temperature a transition to the Bose-Einstein condensation. This concept can explain a lot of experiments such as excess conductivity [22], ultrafast carrier localization into polaronic state [23], Nernst effect as one of the most convincing demonstration for the existence of the preformed pairs [3], tunneling and spectroscopic properties [24], [16].

Here we have shown that the existence of the pseudogap state and non-standard behavior of the isotope coefficient in HTSC materials can be explained on the basis of the electron-phonon interaction without the involvement of any other scenarios [25]-[29].

We witness further discussion on the nature of the pseudogap phase in HTSC materials (see for example [30], [31]). It follows from the foregoing consideration that the pseudogap is a universal effect and should arise as TI bipolarons are formed in a system. The fact that for a long time the occurrence of the pseudogap phase was associated with side effects is caused by that this phase is observed even in ordinary SC [32]-[37], where the occurrence of the pseudogap was explained by a crystallographic disorder or reduced dimensionality which are usually observed in disordered metals.

In this connection recent experiments with MgB$_2$ HTSC seem to be of importance [38]. As distinct from oxide ceramics, MgB$_2$ does not have a magnetic order and, as proponents of an external nature of the pseudogap state suggest, should not have a pseudogap. To exclude any other possibilities concerned with disorder, low-dimensionality effects, etc. in experiment [38] use was made of highly perfect crystals. Experiments made in [38] convincingly demonstrated the availability of the pseudogap state in MgB$_2$ and responsibility of EPI for this state. The results obtained provide good evidence for the TI bipolaron mechanism of the formation of the pseudogap state.

The simple scenario presented in the paper may overlap with the effects associated with spin fluctuations, formation of charge density waves (CDW) and spin density waves (SDW), pair density waves (PDW) and bond density waves (BDW), formation of stripes (for example, a giant isotopic effect caused by EPI was observed in La$_{2-x}$Sr$_x$CuO$_4$ in the vicinity of the temperature of charged stripe ordering when replacing $^{16}$O by $^{18}$O [39]), clusters, other types of interactions, etc. The suggested TI bipolaron mechanism of the pseudogap phase formation and explanation of isotopic effects in HTSC materials on its basis are also important in view of universality of this mechanism. The most of HTSC are spatially inhomogeneous. Therefore, one can think that the translation invariant bipolarons are not applicable there. However, this can be not correct as long as inhomogeneity is not very large. As shown in [40] the crystal defects can capture and destroy

TI- bipolaron only if their potential well is sufficiently large. Otherwise, TI-bipolaron remains stable.

In recent paper by the author [41] it was shown that the TI-bipolaron theory of SC with strong EPI can explain the dependence of an isotope coefficient in HTSC on the critical temperature of SC transition and London penetration depth. For the case of small radius polaron this explanation earlier was done by Alexandrov (for review see [42]). Due to the large value of bipolaron effective mass, the small radius bipolaron theory can explain a lot number of effects in low-temperature SC with strong electron-phonon coupling and is less appropriate for HTSC. The reason why the EPI in HTSC materials is strong lies in the dominant role of $\omega_0$ near the nodal direction even when it has a small value (EPI coupling is infinite if $\omega_0$ is zero in nodal direction).

At present, there are only a few experiments to study the isotope effect in the pseudogap phase. The experiments on the influence of the magnetic field on the isotope coefficient proposed in the paper for the temperature of the transition to the pseudogap phase are new. The agreement of the experimental results with the theoretical predictions would indicate the validity of the assumption of the TI bipolaron mechanism of HTSC.

**Appendix A. Derivation of Formula (13)**

Taking into account the relation $\omega_0 \sim M^{-1/2}$ from (12) we will get:

$$\alpha^* = \frac{1}{2}\frac{d\ln T^*}{d\ln \omega_0} = \frac{\omega_0}{2T^*}\frac{dT^*}{d\omega_0} \tag{A1}$$

It follows from (8) that:

$$\frac{dT^*}{d\omega_0} = æ^{-1} e^W \frac{dW}{d\omega_0}, \tag{A2}$$

$$\frac{dW}{d\omega_0} = \frac{dW}{dz}\frac{dz}{d\omega_0} = \frac{e^{-W}}{1+W}\frac{dz}{d\omega_0}.$$

Taking into account that $dz/d\omega_0 = 2æ/3$ for $dT^*/d\omega_0$ determined by (A2), we will get:

$$\frac{dT^*}{d\omega_0} = \frac{2}{3(1+W)} \tag{A3}$$

It follows from (A1) and (A3) the expression (13) for isotope coefficient $\alpha^*$:

$$\alpha^* = \frac{\omega_0}{3T^*}\frac{1}{1+W(z)}$$